\documentclass[pra,amsfonts,amsmath,twocolumn,superscriptaddress,nofootinbib]{revtex4}

\usepackage{color}
\usepackage{graphicx} 
\usepackage{epstopdf}%

\newcommand {\be} {\begin{equation}} 
\newcommand {\ba}{\begin{eqnarray}} 
\newcommand {\ee} {\end{equation}} 
\newcommand{\ea} {\end{eqnarray}}

\renewcommand{\epsilon}{\varepsilon}

\begin{document}

\title{Constraining off-shell effects using low-energy Compton scattering}

\author{Carl E.\ Carlson}
\affiliation{Department of Physics, College of William and Mary, Williamsburg, VA 23187, USA}

\author{Marc Vanderhaeghen}
\affiliation{Institut f\"ur Kernphysik, Johannes Gutenberg-Universit\"at, D-55099 Mainz, Germany}

\date{\today}

\begin{abstract}
Off-shell effects in proton electromagnetic vertices can be constrained from their effects on known processes.  In particular, parameters in models for the off-shell effects can be determined by fitting to the proton electric and magnetic polarizabilities measured in low-energy Compton scattering.  There has been recent speculation that off-shell effects  contribute enough energy to the muonic hydrogen Lamb shift to explain the discrepancy between muonic and electronic measurements of the proton radius.  We find that the constraints discussed here make the off-shell effects about two orders of magnitude smaller than needed for this purpose.
\end{abstract}

\maketitle

%%%%%%%%%%%%%%%%%%%%%%%%%%%%%%%%%%%%%%%%%%%%

\section{Introduction}			\label{sec:one}

%%%%%%%%%%%%%%%%%%%%%%%%%%%%%%%%%%%%%%%%%%%%

\begin{figure}[bthp]
\bigskip
\begin{center}
\includegraphics[width = 50 mm]{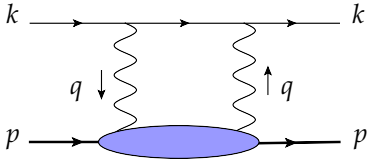}
\caption{The general box diagram for the $\mathcal O(\alpha^5)$ corrections.}
\label{fig:lambbox}
\end{center}
\end{figure}

The $\mathcal O(\alpha^5)$ corrections to the Lamb shift are given by the two-photon exchange box diagram shown in Fig.~\ref{fig:lambbox}.   Part of this diagram was first calculated in a coordinate-space  formalism by Friar~\cite{Friar:1978wv} some time ago, and a complete calculation was first given by Pachucki~\cite{Pachucki:1999zza}.   The calculation is well done using dispersion relations to obtain the real part of the box once the imaginary part is obtained.  The imaginary part is obtained using the optical theorem, which requires only knowledge  of the amplitudes with real and on-shell electron and hadron intermediate states.  The dispersion relation sums and integrates over all kinematically allowed intermediate states, so that elastic and inelastic intermediate states are simultaneously included.  

%%%%%%%%%%%%%%%
%%%%%%%%%%%%%%%

\begin{figure}[bthp]
\begin{center}
\includegraphics[width = 82 mm]{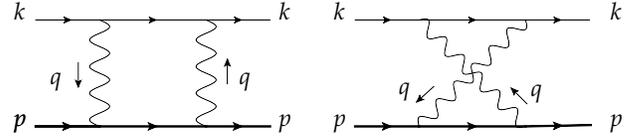}
\caption{The box diagram with proton intermediate states.
}
\label{fig:el}
\end{center}
\end{figure}

The elastic contributions, those with proton intermediate states, are shown in Fig.~\ref{fig:el}.  Again,  the imaginary part of these diagrams gets contributions only with electrons and protons on-shell, in which case the photon-proton vertex can be written completely using only the Dirac and Pauli terms, as
\begin{equation}
\Gamma^\mu = \gamma^\mu F_1(Q^2) 
		+ \frac{i}{2M} \sigma^{\mu\nu}q_\nu F_2(Q^2)
\label{eq:onshell}
\end{equation}
for an incoming photon of momentum $q$.

%%%%%%%%%%%%%%%
%%%%%%%%%%%%%%%

One can entertain the idea of obtaining both the real and imaginary parts of the elastic contributions from a direct calculation of the box diagram (Fig.~\ref{fig:el}).  In this case there will be off-shell intermediate states, and there can be extra terms at the photon-proton vertex, both in the sense of extra Lorentz structures and of the existing form factors depending additionally on how far the proton is off-shell.  The new or modified form factors contain parameters which are initially unknown, and one can and should attempt to constrain them by comparison with existing data.

The current motivation to reexamine the contributions to the Lamb shift, of course, is the discrepancy between the proton radius obtained from the Lamb shift in muonic 
hydrogen~\cite{Pohl:2010zz} 
and the proton radius obtained from electronic experiments, either from electron 
scattering~\cite{Bernauer:2010wm,Zhan:2011ji} 
or from energy splittings in ordinary hydrogen~\cite{Mohr:2008fa}.  The radius is measured by the effect of the finite proton size upon the atomic energy splittings, and if one could find an additional 0.31 meV or 310 $\mu$eV energy shift in the 2S level of muonic hydrogen, either from new physics or from effects that have been overlooked, then all experiments would be reconciled.

Searching for effects that have been overlooked, Miller \textit{et al.}~\cite{Miller:2011yw} 
have directly calculated the elastic box diagram contribution to the muonic hydrogen Lamb shift, using several models for the off-shell contributions to the proton vertices.  They find that, for a choice of parameters, the effect is large enough to explain the discrepancy between the muonic and electronic measurements of the proton radius.

One should here repeat that the dispersive calculation is a complete calculation, and includes all of what would be off-shell contributions in another formalism.  The dispersive calculation depends on being able to obtain the imaginary part of the box from experimental data on elastic and inelastic electron scattering, and this is possible here~\cite{Pachucki:1999zza,Martynenko:2005rc}.  
A recent re-evaluation~\cite{Carlson:2011zd} has both updated this contribution and 
quantified its error resulting from the most recent input. The resulting energy correction on the 2S level in muonic hydrogen yields~: $\Delta E = -36.9 \pm 2.4~\mu$eV. 
Hence there is, strictly speaking, no need for a direct calculation.  However, one could take the view that the direct calculation gives a way to model the non-pole and inelastic parts of the $\alpha^5$ calculation and so gives a thinkable approximation worthy of pursuit.  One would expect the answers to be of the same order of magnitude.  This has not so far proved to be the case, and so it is necessary to understand how the direct calculation of Ref.~\cite{Miller:2011yw}  putatively obtained hundreds of $\mu$eV energy when the dispersive calculation was an order of magnitude smaller.

More generally, one can ask how the parameters that enter the forms chosen for the off-shell photon-proton vertex can be constrained by experimental data on known processes.  In particular proton Compton scattering is well measured.  One notices, of course, that the lower part of the elastic box diagram is just forward off-shell Compton scattering,  so that the off-shell vertices that enter the box diagram also give contributions beyond the Born approximation to Compton scattering.  (The Born approximation is defined as the calculation using only the on-shell vertex, Eq.~(\ref{eq:onshell}).)   The Compton amplitudes, at low energy,  are given in terms of the measured electric and magnetic polarizabilities.  We shall describe some details of how this is done, and subsequently of how the parameters in the several model off-shell vertices chosen in Ref.~\cite{Miller:2011yw} are related to and hence determined by the polarizabilities.  The consequences of the contribution of the box diagrams to the Lamb shift is that the results are about two orders of magnitude smaller than hoped for in~\cite{Miller:2011yw}, and compatible with the dispersive results.

%%%%%%%%%%%%%%%%%%%%%%%%%%%%%%%

\section{Compton scattering and polarizabilities}		\label{sec:basic}

%%%%%%%%%%%%%%%%%%%%%%%%%%%%%%%

Forward Compton scattering can be described by the Compton tensor,
\begin{align}
&T^{\mu\nu}(p,q) = \frac{i}{8\pi M}  \int d^4x	\,e^{iqx}
	\langle p | T j^\mu(x) j^\nu(0) | p \rangle  \nonumber\\[1ex]
	& \quad = \left(-g^{\mu\nu} + \frac{q^\mu q^\nu}{q^2} \right) T_1(\nu,Q^2)
		\nonumber\\
	& \quad + \frac{1}{M^2} \left( p^\mu - \frac{p\cdot q}{q^2} q^\mu \right)
	\left( p^\nu - \frac{p\cdot q}{q^2} q^\nu \right)  T_2(\nu,Q^2)	,
\end{align}
where an average over the proton polarization is implied.
The amplitude for forward Compton scattering is given in terms of the Compton tensor by
\be
{\mathcal M}_{\lambda'_\gamma \lambda_\gamma} 
	=  8\pi M e^2 \,  {\varepsilon}_\mu^*(q,\lambda'_\gamma) 
	\varepsilon_\nu(q,\lambda_\gamma)  T^{\mu\nu}	\,,
\ee
where $\lambda_\gamma$ is a photon polarization.  From the Born terms alone,
\begin{align}
T_1^B(\nu,Q^2) &= \frac{1}{4\pi M} \left\{ 
	\frac{Q^4 (F_1+F_2)^2 }{(Q^2-i\epsilon)^2 - 4M^2 \nu^2} - F_1^2
	\right\}		\,,
					\nonumber\\
T_2^B(\nu,Q^2) &= \frac{M  Q^2}{ \pi  }  
	\frac{F_1^2 + (Q^2/4M^2) F_2^2}
		{(Q^2-i\epsilon)^2 - 4M^2 \nu^2}	\,.
\end{align}

Corrections beyond the Born term can be given in terms of an effective Hamiltonian expansion,  which must be quadratic in the vector potential and also be gauge invariant.  It must also be Lorentz invariant and parity and time reversal invariant. In terms of the fields, the leading form for a spin-averaged proton is, see e.g.~\cite{Babusci:1998ww,Drechsel:2002ar}, given by~:
\be
\mathcal H_{\rm eff} = - \frac{1}{2} 4\pi \alpha_E \vec E^{\,2} 
	- \frac{1}{2} 4\pi \beta_M \vec B^{\,2}	,
\ee
and $\alpha_E$ and $\beta_M$ are the electric and magnetic polarizabilities.  This is also, in the proton rest frame,
\begin{align}
\mathcal H_{\rm eff} = 
	\frac{1}{2} 4\pi \left( \alpha_E + \beta_M \right)
	\frac{p^\alpha   F_{\alpha\nu}     p_\beta F^{\beta\nu} }{M^2}
	- \frac{1}{4} 4\pi \beta_M   F_{\mu\nu} F^{\mu\nu}	,
\end{align}
from which one obtains the leading corrections
\begin{align}				\label{eq:nonborn}
\lim_{\nu,Q^2\to 0}  T_1^{\rm non-Born}(\nu,Q^2) &=  \frac{\nu^2}{e^2} \left( \alpha_E +\beta_M \right)
	+ \frac{Q^2}{e^2} \beta_M	\,,
		\nonumber \\
\lim_{\nu,Q^2\to 0}  T_2^{\rm non-Born}(\nu,Q^2) &=  \frac{Q^2}{e^2} \left( \alpha_E +\beta_M \right)		\,.
\end{align}

The same results follow from the non-relativistic QED formalism, as shown by 
Hill and Paz~\cite{Hill:2011wy}. 
They present explicitly the result for $T_1(0,Q^2)$, which is precisely the Born term 
expanded to $\mathcal O(Q^2)$ plus the above.

%%%%%%%%%%%%%%%%%%%%%%%%%%

\section{Off-shell contributions}   \label{sec:off}

%%%%%%%%%%%%%%%%%%%%%%%%%%

There are many ways to include off-shell effects for the proton.  We will consider specifically the three suggestions of Miller \textit{et al.}~\cite{Miller:2011yw}, 
who change the Dirac part of the vertex function, so that at a vertex with incoming and outgoing proton momenta $p$ and $p'=p+q$ one has
\be
\Gamma^\mu_{\rm Dirac} = \gamma^\mu F_1(Q^2) + F_1(Q^2) F(Q^2) 
	\mathcal O^\mu_{a,b,c}
\ee
with $F(0)=0$ to maintain the proton charge normalization.  Further,
\begin{align}
\mathcal O^\mu_a &= \frac{(p+p')^\mu}{2M}
	\left[ \Lambda_+(p') \frac{\not\! p - M}{M} 
	+  \frac{\not\! p' - M}{M} \Lambda_+(p) \right]	\,,
			\nonumber\\
\mathcal O^\mu_b &= \gamma^\mu \left( 
\frac{p^2-M^2}{M^2} + \frac{{p'}^2-M^2}{M^2} \right)	\,,
			\nonumber\\
\mathcal O^\mu_c &= \Lambda_+(p') \gamma^\mu \frac{\not\! p - M}{M}
	+  \frac{\not\! p' - M}{M} \gamma^\mu \Lambda_+(p)	\,,
\end{align}
for $\Lambda_+(p) = ( \!\not\! p +M)/(2M)$.
The parameterization for $F(Q^2)$ is
\be
F(Q^2) = \frac{\lambda Q^2/b^2}{\big( 1 + Q^2/0.71 {\rm\ GeV}^2 \big)^{1+\xi}}  \,.
\ee
Regarding application to the Lamb shift, the results are not sensitive to $\xi$ and Miller \textit{et al.}~give numerical results for $\xi=0$.  To obtain an energy shift of the correct size to explain the proton charge radius discrepancy, they quote~\cite{Miller:2011yw}~:
\be
\left. \frac{\lambda}{b^2} \right|_{\rm Miller\ et\ al.}
= \left\{	\begin{array}{lr}
\frac{2}{(79 {\rm\ MeV})^2} \approx 320 {\rm\ GeV}^{-2} 
	& {\rm for\ }\mathcal O_a^\mu ,	\\[1.3ex]
\frac{3}{5} \times above & \mathcal O_b^\mu ,	\\[1ex]
-\frac{3}{2} \times above & \mathcal O_c^\mu .	
\end{array}
\right.
\ee

One may now calculate proton Compton scattering using diagrams with a single proton exchange,  using the modified vertices.  One obtains the Born terms, plus terms with an off-shell correction in one vertex and terms with off-shell corrections in both vertices.  For connecting with the polarizabilities, we need only corrections with two and not more powers of the photon momenta.  There is already a factor $Q^2$ in $F(Q^2)$, so terms with two off-shell corrections are dropped.  Similarly the Pauli terms do not enter the corrections since they contain a visible extra factor $q$.   One finds the off-shell corrections
\begin{align}
T_1^{\rm off} &= \left\{ \begin{array}{lc}
0 & \mathcal O_a^\mu , \mathcal O_b^\mu ,	\\[1ex]
\frac{1}{\pi M} F_1^2 F + \mathcal O(q^4) \approx \frac{\lambda}{\pi M b^2} Q^2
	&	\mathcal O_c^\mu ,
\end{array}	\right.
\nonumber \\[2ex]
T_2^{\rm off} &= \left\{ \begin{array}{lc}
\frac{-1}{\pi M} F_1^2 F + \mathcal O(q^4) \approx \frac{-\lambda}{\pi M b^2} Q^2
	& 	\quad \mathcal O_a^\mu ,			\\[1.3ex]
\frac{-2}{\pi M} F_1^2 F + \mathcal O(q^4) \approx\frac{-2\lambda}{\pi Mb^2} Q^2
	&	\quad \mathcal O_b^\mu ,			\\[1.3ex]
0	&	\quad \mathcal O_b^\mu .
\end{array}	\right.
\end{align}

These corrections show no falloff with $\nu$.  This is because the $\nu$ dependence that would come from the denominator of the proton propagator is canceled by numerator terms in the off-shell corrections.  Hence a dispersion relation in $\nu$ for $T_2(\nu,Q^2)$ would require a subtraction, contrary to the usual practice.

A model applied to Compton scattering should yield results that match the results from the gauge invariant effective expansion involving the electric and magnetic polarizabilities, Eq.~(\ref{eq:nonborn}).  The forms in Ref.~\cite{Miller:2011yw} do not.  Some completion of them must, but for now we will simply estimate the parameter in the Ref.~\cite{Miller:2011yw} models by matching the corresponding terms with non-zero $Q^2$ dependences.  This gives $\lambda/b^2$ from the polarizability data as
\be
\left. \frac{\lambda}{b^2} \right|_{\rm pol.\ data}
= \left\{	\begin{array}{clr}
-\frac{M}{4\alpha} (\alpha_E+\beta_M ) & \approx -5.8 {\rm\ GeV}^{-2} 
	& \ \  \mathcal O_a^\mu ,	\\[1.3ex]
-\frac{M}{8\alpha} (\alpha_E+\beta_M ) & \approx -2.9 {\rm\ GeV}^{-2} 
	& \mathcal O_b^\mu ,	\\[1.3ex]
\frac{M}{4\alpha} \beta_M & \approx + 1.4 {\rm\ GeV}^{-2} 
	& \mathcal O_c^\mu .	
\end{array}
\right.
\ee
using $\alpha_E + \beta_M = 0.00139(4)$ fm$^3$ from 
Drechsel \textit{et al.}~\cite{Drechsel:2002ar} or Schumacher~\cite{Schumacher:2005an}, 
and $\beta_M = 0.00034(12)$ fm$^3$ from Beane \textit{et al.}~\cite{Beane:2004ra} (see also~\cite{Schumacher:2005an} and~\cite{Lensky:2009uv}).  

Hence the values for $(\lambda/b^2)$, which enter linearly in the expressions for the contributions to the Lamb shift energy, are as obtained from the Compton scattering data $55, 67,$ and $340$ times smaller than the values needed to explain the proton radius discrepancy, for models $a$, $b$, and $c$, respectively.  The sign is also incorrect.

%%%%%%%%%%%%%%%%%%%%%%%%%%

\section{Discussion}				\label{sec:end}

%%%%%%%%%%%%%%%%%%%%%%%%%%

The dispersive analysis of the two-photon structure dependent contributions to the Lamb shift, represented by the box diagram of Fig.~\ref{fig:lambbox}, gives the full, complete, and correct result.  It allows one to obtain the full Compton amplitude in terms of its imaginary part plus a subtraction constant. The imaginary part is obtainable from experimental electron-proton scattering data, and the subtraction constant is proportional to the magnetic polarizability obtained from low-energy Compton scattering on the proton. 
  The dispersion relation method requires no separate investigation of off-shell proton contributions.  Hence, from the dispersive calculation one obtains a result as complete as the relevant experimental data will allow, and one can estimate reliably limits on contributions from experimentally unexplored regions.

While the off-shell contributions need not be separately included,  one might argue that they could be used as a model of the non-nucleon-pole or non-Born contributions due to the strong interactions.  Indeed, one can view the present discussion as an opportunity to show how to estimate or constrain the parameters involved in any model for the interactions of nucleons off-shell.

Miller \textit{et al.}~\cite{Miller:2011yw} have presented a model of the off-shell contributions, albeit one with the feature that it leads to a $T_2$ Compton amplitude that required a subtraction were it to be calculated dispersively.   This stands in contrast with the standard analysis of its high energy behavior.  One may work with it anyway, but one should compare the resulting $T_1$ and $T_2$ to known expansions of the Compton amplitudes beyond the pole terms, which are given at low energy and momentum in terms of the electric and magnetic polarizabilities, $\alpha_E$ and $\beta_M$.  This forces a serious constraint upon any parameterization of off-shell behavior.  This constraint, proportional to the measured $\alpha_E$ and $\beta_M$, leads to much smaller values of the crucial parameter than desired in~\cite{Miller:2011yw}.  The decrease in the magnitude of the estimated contribution to the Lamb shift is by a factor of order $50$ to $300$, depending on the specifics of the model.

%%%%%%%%%%%%%%%%%%%%%%%%%%%%%%%

\begin{acknowledgments}

CEC thanks the National Science Foundation for support under Grant PHY-0855618, and thanks the Helmholtz Gemeinschaft in Mainz for its hospitality.  We also thank Gerry Miller and Gil Paz for discussions.

\end{acknowledgments}

%%%%%%%%%%%%%%%%%%%%%%%%%%%%%%%

\end{document}